\documentstyle[preprint,aps,prl]{revtex}
\begin{document}
\draft

\begin{center}
{\bf Rigorous Density Functional Theory for\\
Inhomogeneous Bose-Condensed Fluids}\\[1cm]

A. Griffin\\
{\it Dipartimento di Fisica, Universit\`{a} di Trento,
I-38050 Povo, Italy\\
and\\
Department of Physics, University of Toronto,
Toronto, Ontario, Canada M5S 1A7} \footnote{permanent
address}

\vspace{2cm}
To appear in the special Brockhouse Issue\\
of the Can.\ Journ.\ Phys., 1995.\\

\vspace{2cm}

Phone: 416-978-5199/7135\\
Fax: 416-978-2537\\
e-mail: griffin@physics.utoronto.ca
\end{center}

\newpage

\begin{abstract}
The density functional theory originally developed by Hohenberg,
Kohn
and Sham provides a rigorous conceptual framework for dealing with
inhomogeneous interacting Fermi systems. We extend this approach to
deal with inhomogeneous interacting Bose-condensed systems,
limiting
this presentation to setting up the formalism to deal with ground
state $(T=0)$ properties. The key
new feature is that one must deal with energy functionals of both
the
local density $n({\bf r})$ and the local complex macroscopic
wavefunction $\Phi ({\bf r})$ associated with the Bose
broken-symmetry
(the local
condensate density is $n_{c}({\bf r}) = \vert \Phi ({\bf r}) \vert
^{2}$). Implementing the Kohn-Sham scheme, we reduce the problem to
a gas of weakly-interacting Bosons moving in self-consistent
diagonal
and off-diagonal one-body potentials.  Our formalism should provide
the basis for studies of the surface properties of liquid $^4$He as
well as the properties of Bose-condensed atomic gases trapped in
external potentials.
\end{abstract}

\vskip 0.1 true in
\newpage

\section{Introduction}

This article develops a density functional formalism for dealing
with
superfluid $^4$He, with the ultimate goal of understanding the role
of Bose-condensation in the low density surface region. The use of
neutron scattering to study the atomic structure and dynamics of
superfluid $^4$He has been an ongoing research effort at Chalk
River
since the early fifties. This world famous programme grew out of
and was nurtured by the atmosphere which Bert Brockhouse helped to
create. On a more personal note, I have always enjoyed being
invited
to give seminars at McMaster, partly for the stimulating
discussions
with research colleagues but also because I knew that Bert would be
in the audience and after it was over, he would come over and give
some encouraging comments. It is with great pleasure that I
dedicate
this article to Prof. Bert Brockhouse.

At a phenomenological level, the low temperature properties of bulk
superfluid $^4$He are well understood in terms of Landau's picture
of a weakly-interacting quasiparticle gas of phonons and rotons. At
a more microscopic (atomistic) level, current discussions of
superfluid $^4$He can be divided into two broad classes. One
of these is the field-theoretic analysis, which is built on the
fundamental role of Bose-broken symmetry. The superfluid phase
coincides with the appearance of a macroscopic
wavefunction given by the finite expectation value of the field
operator $\Phi ({\bf r}) = \langle \hat \psi ({\bf r}) \rangle$
\protect\cite{pch-pcm65}.
This
approach goes back to the pioneering work of Bogoliubov
\protect\cite{nnb47}
and
Penrose \protect\cite{op51}, but was first formulated in a
systematic
way by Beliaev \protect\cite{stb58} and Hugenholtz and Pines
\protect\cite{newref5}.
The current status of theories of superfluid $^4$He based on the
key
role of the
Bose order parameter $\Phi ({\bf r})$ is reviewed in
a recent book ~\protect\cite{ag93}.

A second kind of microscopic theory is built on the use of
variational many-body wavefunctions for the ground state and the
low-energy excited states of liquid $^4$He. Such many-body
wavefunctions were first introduced by Bijl \protect\cite{ab40} and
later by
Feynman \protect\cite{rpf54}, and in their
current form have become very sophisticated (the correlated basis
function approach \protect\cite{ef69}). However, while such ground
state many-particle
wavefunctions (of the Jastrow-Feenberg type, for example) lead to
very good estimates of the condensate fraction (about 10\% at
$T=0$), the role of a Bose order parameter is not exhibited very
explicitly (see chapter 9 of \protect\cite{ag93}).

The motivation of the present paper is to set up a formalism which
can deal with the surface
properties of superfluid $^4$He, taking Bose-condensation into
account from the beginning.
Formally
 this means we need a theory of spatially {\it inhomogeneous}
Bose-condensed system. In the last decade, there has developed a
considerable literature on superfluid
$^4$He with free surfaces (including droplets and films) based on
generalizations of the above mentioned correlated basis function
approach \protect\cite{ek-etc94}. However, as with the case of bulk
liquid
$^4$He, such treatments of surfaces give little (if any)
emphasis to the role of
Bose-broken symmetry or Bose-Einstein condensation.
On the other hand, the many-body wavefunctions which have
been
developed \protect\cite{ccc-mc73} to describe the free surface of
liquid $^4$He at $T=0$
appear to give realistic estimates of the density profile $n({\bf
r})$ in
the
surface region.  From these, one finds accurate values of the
binding
energy of $^3$He and
spin-polarized H atoms bound to the surface of superfluid $^4$He
\protect\cite{ibm-doe79}.

Such correlated basis function approaches involve a very
heavy computational effort when dealing with inhomogeneous systems.
An alternative, much simpler theory has developed for free
surfaces of liquid $^4$He
based on a density functional theory \protect\cite{ce-wfs-etc75}.
This approach is loosely inspired by the very
successful
theory of inhomogeneous interacting electron systems and
it is useful to
recall some aspects of the latter theory. Almost 30 years ago,
Hohenberg and Kohn (HK) gave a rigorous formulation
\protect\cite{ph-wk65} of the ground
state properties of interacting Fermi systems, in terms of a
functional of the local density $n({\bf r}) = \langle \hat
\psi^{\dagger}
({\bf r}) \hat \psi ({\bf r}) \rangle$. Kohn and Sham
\protect\cite{wk-ljs65} (KS) used the
two exact HK theorems to implement the HK formalism in terms of
finding the single-particle energies and eigenstates of free
Fermions moving in
an appropriately defined self-consistent fields. This
Hohenberg-Kohn-Sham (HKS) density functional formalism is now the
accepted way of
dealing with inhomogeneous Fermi systems, building on our extensive
knowledge
of the ground state properties of homogeneous interacting Fermi
systems. The HKS approach has been generalized to deal with normal
systems at finite temperatures \protect\cite{dm-inc} as well as BCS
superconductors \protect\cite{lno-ekg-wk88}. In the latter case,
one
works with functionals of
the local density $n({\bf r})$ and the local anomalous (or
off-diagonal) density $\Delta ({\bf r}) \equiv \langle \hat
\psi_{\uparrow} ({\bf r})\hat \psi_{\downarrow} ({\bf r}) \rangle$
describing spin-singlet Cooper pairs. Finally, a density functional
formalism
has
been developed for dealing with time-dependent quantities
(excited-states and linear response functions) of both normal
\protect\cite{er-ekg84} and
superconducting Fermi systems \protect\cite{ojw-rk-ekg94}.

As we have noted, recent density functional theories of superfluid
$^4$He with
surfaces make analogies to the above HKS theory of inhomogeneous
Fermi
systems. However, there has apparently never been a careful study
of
how to
use
the HKS ideas to give a rigorous basis to a theory of inhomogeneous
Bose-condensed liquids, analogous to what has been done
for inhomogeneous BCS superconductors \protect\cite{lno-ekg-wk88}.
In particular, all current
$T=0$ density functional theories of superfluid $^4$He
\protect\cite{ce-wfs-etc75} simply
assume
that the energy functional only depends on the local density
$n({\bf
r})$. There is never any reference
to the possibly equally important role that the local condensate
density
$n_{c}({\bf r})$ might play, or more generally, the local
macroscopic
wavefunction $\Phi ({\bf r}) = \langle \hat \psi ({\bf r}) \rangle
= \sqrt{n_{c}({\bf r})} e^{iS({\bf r})}$.
Needless to say, it is the finite value of the order parameter
$\Phi
({\bf r})$ which characterizes the superfluid phase below
$T_{\lambda} = 2.17 K$. To avoid confusion, we note that the energy
in density functional
theories of liquid Helium (see third paper of ref.
\protect\cite{ce-wfs-etc75}) is often taken to be a functional of
the
variable $\Psi ({\bf r}) \equiv \sqrt{n({\bf r})}e^{iS({\bf r})}$,
which
involves the {\it total} local density. While this variable is
sometimes
referred to as a ``macroscopic wavefunction'', it is clearly
unrelated to the Bose order parameter $\Phi ({\bf r})$ we have
introduced above.
The insufficiency of theories based on
functionals of only $n({\bf r})$ has been made especially
obvious in recent work \protect\cite{ag-ss-sub} which points out
that
the low density
surface
region of liquid $^4$He corresponds to a dilute inhomogeneous
Bose gas
with 100\% Bose condensation (at $T=0$). In
this surface region, the key function is
$\Phi ({\bf r})$,
with the local density being determined by it, namely
$n({\bf r}) = n_{c}({\bf r}) \equiv \vert \Phi ({\bf r}) \vert
^{2}$.

In the present paper, we formulate a density functional theory of
the
HKS kind for the ground state properties of an inhomogeneous
interacting
Bose-condensed fluid. The key new element in our analysis
(following
the analogous case of BCS superconductors
\protect\cite{lno-ekg-wk88})
is the
realization that one must work with functionals of both $n({\bf
r})$
and the (complex) order parameter $\Phi ({\bf r})$. The key
theorems
of HK and the methods of proof are, of course, valid for Bose as
well
as Fermi statistics. For brevity, we shall only sketch these
arguments when they involve the identical steps as in the density
functional treatment of BCS superconductors. The present analysis
of ground state properties can be extended to finite
temperatures (free energies) following the analogous discussion for
BCS
superconductors \protect\cite{lno-ekg-wk88}.  This will be reported
elsewhere.

What the present paper accomplishes is to give a formally exact
scheme for dealing with inhomogeneous Bose-condensed fluids
which should ultimately provide a platform for specific
calculations.
In applying the present formalism, one must introduce
approximations for the correlation energy functionals. This is a
separate
question, with specific problems associated with the anomalous
long-range
correlations in Bose-condensed systems \protect\cite{yan-lrp94},
and
is not treated here.

In this paper, we mainly use the surface region of superfluid
$^4$He as an example of an inhomogeneous
Bose-condensed system. Very recently, Bose-condensation has been
finally achieved \cite{newref22} in a dilute gas of $^{87}$Rb atoms
below 200 nK, using laser and evaporative cooling.  This gas was
trapped in a harmonic potential well and as a result, both $n({\bf
r})$ and $n_c({\bf r})$ are highly inhomogeneous. Our present
formalism gives a natural basis for generalizing the currently
available
Hartree-Fock-Bogoliubov approximations \cite{newref23} for such
inhomogeneous weakly interacting Bose-condensed gases.
\section{Hohenberg-Kohn Formalism}

Our starting Hamiltonian is defined as (compare with
\protect\cite{lno-ekg-wk88})
\begin{eqnarray}
\hat H_{v,\eta}
&\equiv&
\int d{\bf r} \hat \psi^{\dagger}({\bf r})\
\left[ -\frac{\nabla ^{2}}{2m} - \mu \right] \hat \psi ({\bf r})
\nonumber\\
&&
+ \frac{1}{2} \int d{\bf r} \int d{\bf r}^{\prime}
\hat \psi ^{\dagger}({\bf r}) \hat \psi ^{\dagger}({\bf
r}^{\prime})
v_{2}({\bf r} - {\bf r}^{\prime}) \hat \psi ({\bf r}^{\prime})
\hat \psi ({\bf r})
\nonumber\\
&&
+ \int d{\bf r} v({\bf r}) \hat \psi ^{\dagger} ({\bf r}) \hat \psi
({\bf r})
\nonumber\\
&&
+ \int d{\bf r} [\eta ({\bf r}) \hat \psi ^{\dagger} ({\bf r}) +
\eta ^{*} ({\bf r}) \hat \psi ({\bf r})]\label{eq1}\\
&\equiv&
\hat H_{0} - \mu \hat N + \hat V_{2} + \hat V_{1} +
\hat V_{SB}.\label{eq2}
\end{eqnarray}
Throughout the analysis, the two-particle interaction
$v_{2}({\bf r} - {\bf r}^{\prime})$
is assumed to be fixed. Thus $\hat H_{v,\eta}$ in (\ref{eq1})
depends only on: (a) the choice of the external diagonal
single-particle potential $v({\bf r})$, which couples to the
density;
and
(b) the external off-diagonal (or symmetry-breaking) potential
$\eta
({\bf r})$, which couples to the field operators $\hat \psi ({\bf
r})$ and $\hat \psi ^{\dagger} ({\bf r})$. The first step in the HK
approach is to note that the ground state $\vert \Psi \rangle$ of
$\hat H_{v,\eta}$ is a functional of these external fields $v({\bf
r})$ and $\eta ({\bf r})$, and hence so are the following
groundstate expectation
values:
\begin{eqnarray}
n({\bf r})&\equiv&\langle \Psi \vert \hat \psi ^{\dagger} ({\bf r})
\hat \psi ({\bf r}) \vert \Psi \rangle
\nonumber\\
\Phi ({\bf r})&\equiv&\langle \Psi \vert \hat \psi ({\bf r}) \vert
\Psi \rangle
\nonumber\\
\Phi ^{*} ({\bf r})&\equiv&\langle \Psi \vert \hat \psi ^{\dagger}
({\bf r}) \vert \Psi \rangle .\label{eq3}
\end{eqnarray}
We note that because of the symmetry-breaking field
$\hat V _{SB}$ in
(\protect\ref{eq1}), the ground state eigenstate $\vert \Psi
\rangle$ of $\hat H_{v,\eta}$ allows $\Phi ({\bf r})$ and $\Phi
^{*}
({\bf r})$ to be finite.  Introducing a
symmetry-breaking (off-diagonal) perturbation as in (\ref{eq1}) is
the standard method \cite{pch-pcm65,stb58} of dealing with the
appearance of Bose-condensation in an interacting system. It
involves treating the condensate as a ``reservoir'' of atoms and
thus leads to number non-conservation and to a groundstate where
$\Phi({\bf r})$ can be finite. The underlying physics of this
broken symmetry is the same
for both homogeneous and inhomogeneous systems and is discussed
most clearly in Ref.\cite{pch-pcm65}.  We recall that the essential
physics for dealing with BCS superconductors \cite{lno-ekg-wk88}
also involves such symmetry-breaking number non-conserving states.

The second step of HK is to note that, using their famous {\it
reductio
ad
absurdum} argument, that up to additive constants, one can prove
\begin{equation}
v({\bf r})\ {\rm and}\ \eta ({\bf r})\ {\rm are}\ unique\
{\rm functionals\ of}\ n({\bf r})\ {\rm and}\ \Phi ({\bf r}).
\label{eq4}
\end{equation}
Since $v$ and $\eta$ fix $\hat H_{v,\eta}$ and thus also $\vert
\Psi
\rangle$, we can finally conclude that
\begin{equation}
\vert \Psi \rangle\ {\rm is\ a}\ unique\ {\rm functional\ of}\
n({\bf
r})\
{\rm and}\ \Phi ({\bf r}).
\label{eq5}
\end{equation}
In turn, it follows from (\protect\ref{eq5}) that the expectation
value $\langle \Psi
\vert \hat H_{0} - \mu \hat N + \hat V_{2} \vert \Psi \rangle$ is
a
universal functional of $n({\bf r})$ and $\Phi ({\bf r})$.
That is to say, there is
no explicit dependence of $\vert \Psi \rangle$ on the specific
forms
assumed for $v({\bf
r})$
and $\eta ({\bf r})$, since these can be expressed as functionals
of
$n({\bf r})$ and $\Phi ({\bf r})$, as stated in
(\protect\ref{eq4}).
Summarizing this train of
argument, we conclude that
\begin{equation}
F [n({\bf r}),\Phi({\bf r})] \equiv \langle \Psi \vert \hat H_{0}
- \mu \hat N + \hat V_{2} \vert \Psi \rangle
\label{eq6}
\end{equation}
is a {\it universal} functional of $n$ and $\Phi$, valid for any
number
of
particles and any external potentials $v({\bf r})$ and $\eta
({\bf r})$. As in the HK formalism for Fermi systems, the
universal functional
$F [n,\Phi]$
will play a central role in our subsequent analysis. A key
problem, of course, will be to find some appropriate approximation
to
this
universal functional $F [n,\Phi]$ in superfluid $^4$He.

Following HK, it is useful to define, for given $v$ and $\eta$
potentials, the energy
functional
\begin{eqnarray}
E_{v,\eta} [n,\Phi]&\equiv&F[n,\Phi] + \int d{\bf r} v({\bf r})
n({\bf r})
\nonumber\\
&&
+ \int d{\bf r} [\eta ({\bf r}) \Phi ^{*} ({\bf r}) + \eta ^{*}
({\bf r}) \Phi ({\bf r})],
\label{eq7}
\end{eqnarray}
where the densities $n({\bf r})$, $\Phi ({\bf r})$ and $\Phi ^{*}
({\bf r})$ are defined in (\protect\ref{eq3}). For simplicity of
notation, we shall generally show functionals as depending
only on $\Phi$,
but
in fact, they depend on {\it both} $\Phi$ and $\Phi ^{*}$.
Following HK,
we have a variational principle, i.e., one can prove that
$E_{v,\eta}
[n,\Phi]$ is a {\it minimum} at the correct values of the densities
$n({\bf r})$ and $\Phi ({\bf r})$ produced by the external
potentials
$v({\bf r})$ and $\eta ({\bf r})$, i.e.,
\begin{equation}
\frac{\delta E_{v,\eta} [n,\Phi]} {\delta n({\bf r})} = 0\ ,\
\frac{\delta E_{v,\eta} [n,\Phi]} {\delta \Phi ({\bf r})} = 0.
\label{eq8}
\end{equation}

\section{Kohn-Sham Procedure}

Following Kohn and Sham \protect\cite{wk-ljs65}, there is a clever
way of finding the
correct
values of $n({\bf r})$ and $\Phi ({\bf r})$ (which, according to
(\protect\ref{eq8}), minimize
$E_{v,\eta} [n,\Phi]$) by solving a
simpler auxiliary problem for which the HK theorems in Section II
are
also valid. To understand the logic of the KS procedure in the
context of our present problem, let us consider an auxiliary system
Hamiltonian defined by
\begin{eqnarray}
\hat H^{s}_{v_{{\it s}}, \eta _{{\it s}}} &\equiv&
\int d{\bf r} \hat \psi ^{\dagger} ({\bf r})
\left[-\frac{\nabla ^{2}}{2m} - \mu \right] \hat \psi ({\bf r})
\nonumber\\
&&
+ \hat V_{s} [\hat \psi ^{\dagger}, \hat \psi ] +
\int d{\bf r} v_{s} ({\bf r}) \hat \psi ^{\dagger} ({\bf r})
\hat \psi ({\bf r})
\nonumber\\
&&
+ \int d{\bf r} [\eta _{s} ({\bf r}) \hat \psi ^{\dagger} ({\bf r})
+ \eta^{*}_{s} ({\bf r}) \hat \psi ({\bf r})],
\label{eq9}
\end{eqnarray}
where the interaction $\hat V_{s}$ is a part of $\hat V_{2}$ in
(\protect\ref{eq1}) and (\protect\ref{eq2}), to be specified later.
All the HK results of Section II apply to (\protect\ref{eq9}). In
particular, if we denote the ground state of $\hat H^{s}_{v_{{\it
s}}, \eta _{{\it s}}}$ as $\vert \Psi _{s} \rangle$, then the
densities
\begin{eqnarray}
n_{s} ({\bf r}) &\equiv&
\langle \Psi_{s} \vert \hat \psi ^{\dagger} ({\bf r}) \hat \psi
({\bf
r}) \vert \Psi _{s} \rangle
\nonumber\\
\Phi_{s} ({\bf r}) &\equiv&
\langle \Psi_{s} \vert \hat \psi ({\bf r}) \vert \Psi_{s} \rangle
\label{eq10}
\end{eqnarray}
are unique functionals of the external fields $v_{s}({\bf r})$ and
$\eta_{s}({\bf r})$. In turn, $v_{s}$ and $\eta_{s}$ and hence
$\vert
\Psi_{s} \rangle$ can be shown to be unique functionals of
$n_{s}({\bf r})$ and $\Phi_{s}({\bf r})$. Thus we conclude that
\begin{equation}
F_{s} [n_{s} ({\bf r}), \Phi_{s}({\bf r})] \equiv
\langle \Psi_{s} \vert \hat H_{o} - \mu \hat N + \hat V_{s}
\vert \Psi_{s} \rangle
\label{eq11}
\end{equation}
is a universal functional of $n_{s} ({\bf r})$ and
$\Phi_{s}({\bf r})$. Finally, we can define an energy functional of
this auxiliary system
\begin{equation}
E^{s}_{v_{{\it s}}, \eta _{{\it s}}} [n,\Phi] \equiv
F_{s} [n,\Phi] + \int d{\bf r} v_{s}({\bf r}) n({\bf r})
+ \int d{\bf r} [\eta_{s}({\bf r}) \Phi^{*}({\bf r})
+ \eta^{*}_{s} ({\bf r}) \Phi ({\bf r})],
\label{eq12}
\end{equation}
which will be minimized by the correct values,
$n({\bf r}) = n_{s}({\bf r})$ and
$\Phi ({\bf r}) = \Phi_{s}({\bf r})$, for this system.

The whole point of introducing this auxiliary system  defined by
(\protect\ref{eq9}) is that:

\begin{enumerate}
\item [\rm (a)] It will be easier to solve than the actual system
described by (\protect\ref{eq1}).
\item [\rm (b)] By a judicious choice of the external fields $v_{s}
({\bf r})$ and $\eta _{s}({\bf r})$, the densities $n_{s}({\bf r})$
and
$\Phi _{s}({\bf r})$ of this auxiliary problem can be made to be
{\it identical} to those of the real system. Since $F[n({\bf r}),
\Phi
({\bf r})]$ in (\protect\ref{eq6}) is a universal functional of
only
$n({\bf r})$ and $\Phi ({\bf r})$, this means that we can evaluate
it using results for $n({\bf r})$ and $\Phi ({\bf r})$ obtained
from solving the auxiliary system.
\end{enumerate}

In applying the KS procedure to superconductors
\cite{lno-ekg-wk88},
one uses a non-interacting gas of Fermions moving in
external one-body and pair potentials as the auxiliary model
system.
In our interacting Bose system, this would correspond to setting
$V_{s}[\hat \psi ^{\dagger}, \hat \psi]$ in (\protect\ref{eq9}) to
zero. The problem with this choice is that
for a  non-interacting Bose gas moving in
given
external potentials $v_{s}$ and $\eta _{s}$, and at $T=0$, one
has complete Bose-Einstein condensation (see discussion
after (\ref{neweq35}) for more details). This implies
that $n({\bf r})$ and  $n_{c}({\bf r}) \equiv \vert \Phi ({\bf r})
\vert
^{2}$ are equal,
even though we know that in any interacting Bose system, the local
condensate density $n_{c}({\bf r})$ is
less than the local total
density $n({\bf r})$. This problem is
not addressed
in density functional  theories \protect\cite{ce-wfs-etc75} of
superfluid $^4$He based on functionals of only the density $n({\bf
r})$. We recall that such
theories usually start with the kinetic energy of an inhomogeneous
non-interacting Bose gas with a density profile $n({\bf r})$
identical to the fully-interacting system. Such a kinetic energy
functional implies that $n_c({\bf r})=n({\bf r})$, which
would not appear to be
a very good starting point for describing superfluid $^4$He.

In order to define our auxiliary Bose system in
(\protect\ref{eq9}),
we first introduce the usual decomposition of Bose quantum field operators
\protect\cite{ag93}
\begin{eqnarray}
\hat \psi ({\bf r})
&=& \Phi ({\bf r}) + \tilde \psi ({\bf r})
\nonumber\\
\hat \psi ^{\dagger} ({\bf r})
&=& \Phi ^{*} ({\bf r}) + \tilde \psi ^{\dagger} ({\bf r}),
\label{eq13}
\end{eqnarray}
where $\Phi ({\bf r})$ is defined in (\protect\ref{eq3}). The
non-condensate field operators $\tilde \psi ({\bf r})$ and $\tilde
\psi
^{\dagger} ({\bf r})$ satisfy Bose commutation relations. Using
(\protect\ref{eq13}), the two-particle interaction $\hat V_{2}$ in
(\protect\ref{eq1}) can be rewritten as
\begin{eqnarray}
\hat V_{2}
&=&
\frac{1}{2} \int d{\bf r} \int d{\bf r}^{\prime} v_{2}
({\bf r} - {\bf r}^{\prime}) \vert \Phi ({\bf r}) \vert ^{2}
\vert \Phi ({\bf r}^{\prime}) \vert ^{2}
\nonumber\\
&&
+ \int d{\bf r} \int d{\bf r}^{\prime} v_{2}
({\bf r} - {\bf r}^{\prime}) \vert \Phi ({\bf r}) \vert ^{2}
\Phi ({\bf r}^{\prime}) \tilde \psi ^{\dagger} ({\bf r}^{\prime})
\nonumber\\
&&
+ \int d{\bf r} \int d{\bf r}^{\prime} v_{2}
({\bf r} - {\bf r}^{\prime}) \vert \Phi ({\bf r}) \vert ^{2}
\Phi ^{*} ({\bf r}^{\prime}) \tilde \psi ({\bf r}^{\prime})
\nonumber\\
&&
+ \int d{\bf r} \int d{\bf r}^{\prime} v_{2}
({\bf r} - {\bf r}^{\prime})
\left[ \tilde \psi ^{\dagger} ({\bf r}) \tilde \psi ({\bf r})
\vert \Phi ({\bf r}^{\prime}) \vert ^{2}
+ \tilde \psi ^{\dagger} ({\bf r}) \tilde \psi ({\bf r}^{\prime})
\Phi ^{*} ({\bf r}^{\prime}) \Phi ({\bf r}) \right.
\nonumber\\
&&
\left. + \frac{1}{2} \tilde \psi ^{\dagger} ({\bf r}) \tilde
\psi ^{\dagger}({\bf r}^{\prime}) \Phi ({\bf r}^{\prime}) \Phi
({\bf
r})
+ \frac{1}{2} \tilde \psi ({\bf r}) \tilde \psi ({\bf r}^{\prime})
\Phi ^{*} ({\bf r}^{\prime}) \Phi ^{*} ({\bf r}) \right]
\nonumber\\
&&
+ \int d{\bf r} \int d{\bf r}^{\prime} v_{2}
({\bf r} - {\bf r}^{\prime})
\left[ \tilde \psi ^{\dagger} ({\bf r}^{\prime}) \tilde \psi
({\bf r}^{\prime}) \tilde \psi ({\bf r}) \Phi ^{*} ({\bf r})
+ \tilde \psi ^{\dagger} ({\bf r}^{\prime}) \tilde \psi ^{\dagger}
({\bf r}) \tilde \psi ({\bf r}^{\prime}) \Phi ({\bf r})\right]
\nonumber\\
&&
+ \frac{1}{2} \int d{\bf r} \int d{\bf r}^{\prime} v_{2}
({\bf r} - {\bf r}^{\prime})
\tilde \psi ^{\dagger} ({\bf r}) \tilde \psi ^{\dagger}
({\bf r}^{\prime}) \tilde \psi ({\bf r}^{\prime})
\tilde \psi ({\bf r}).
\label{eq14}
\end{eqnarray}
If all the atoms were Bose-condensed, only the {\it first} term in
(\protect\ref{eq14}) would be important. If the system is not
Bose-condensed, then only the {\it last} term in
(\protect\ref{eq14}) is
present. In the well-known Bogoliubov approximation
\cite{alf-jdw71,alf72} for a dilute,
weakly interacting gas, in which almost all the atoms are
Bose-condensed,
one only keeps terms up to quadratic in the non-condensate field
operators
$\tilde \psi$ and $\tilde \psi ^{\dagger}$ (since it is assumed
that,
in an average sense, $\tilde \psi \ll \Phi$). That is to say, the
last two terms in (\protect\ref{eq14}) are higher order and hence
omitted. A feature of this
Bogoliubov approximation is that the resulting Hamiltonian can be
diagonalized exactly (see below).

We now {\em define} what we shall call (for want of a better term)
the exchange-correlation
energy
functional $F_{xc} [n,\Phi]$ by writing (\protect\ref{eq6}) in the
form
\begin{equation}
F[n,\Phi] = F_{s}[n,\Phi]
+ \frac{1}{2} \int d{\bf r} \int d{\bf r}^{\prime} v_{2}
({\bf r} - {\bf r}^{\prime}) n({\bf r}) n({\bf r}^{\prime})
+ F_{xc}[n,\Phi],
\label{eq15}
\end{equation}
where $F_{s}[n,\Phi]$ is the energy functional of the auxiliary
system {\it defined} by (\protect\ref{eq11}) with the interaction
\begin{eqnarray}
\hat V_{s}
&=&
\int d{\bf r} \int d{\bf r}^{\prime} v_{2}
({\bf r} - {\bf r}^{\prime})
\vert \Phi ({\bf r}) \vert ^{2}
[\Phi ({\bf r}^{\prime}) \tilde \psi ^{\dagger} ({\bf r}^{\prime})
+ \Phi ^{*} ({\bf r}^{\prime}) \tilde \psi ({\bf r}^{\prime})]
\nonumber\\
&&
+ \frac{1}{2} \int d{\bf r} \int d{\bf r}^{\prime} v_{2}
({\bf r} - {\bf r}^{\prime})
[2 \Phi ^{*} ({\bf r}^{\prime}) \Phi ({\bf r}) \tilde \psi
^{\dagger}
({\bf r}) \tilde \psi ({\bf r}^{\prime})
\nonumber\\
&&
+ \Phi ({\bf r}^{\prime}) \Phi ({\bf r}) \tilde \psi ^{\dagger}
({\bf r}) \tilde \psi ^{\dagger} ({\bf r}^{\prime})
+ \Phi ^{*} ({\bf r}^{\prime}) \Phi ^{*} ({\bf r})
\tilde \psi ({\bf r}) \tilde \psi ({\bf r}^{\prime})],
\label{eq16}
\end{eqnarray}
and subject to potentials $v_{s}({\bf r})$ and $\eta _{s}({\bf r})$
[see (\protect\ref{eq12})] chosen such that the density $n_{s}({\bf
r})$ and order parameter $\Phi _{s}({\bf r})$ are {\it identical}
to
those of the full system. As usual \protect\cite{ph-wk65,wk-ljs65},
it is useful to separate out
the
total Hartree energy contribution as we have done in
(\protect\ref{eq15}). Writing
this
contribution out more explicitly for a Bose-condensed system, we
have
\begin{equation}
\langle \hat V _{H} \rangle =
\frac{1}{2} \int d{\bf r} \int d{\bf r}^{\prime} v_{2}
({\bf r} - {\bf r}^{\prime})
[\vert \Phi ({\bf r}) \vert ^{2} \vert \Phi ({\bf r}^{\prime})
\vert
^{2} + 2 \vert \Phi ({\bf r}) \vert ^{2} \tilde n ({\bf
r}^{\prime})
+ \tilde n ({\bf r}) \tilde n ({\bf r}^{\prime})],
\label{eq17}
\end{equation}
where the non-condensate local density is defined by
\begin{equation}
\tilde n ({\bf r}) \equiv \langle \tilde \psi ^{\dagger} ({\bf r})
\tilde \psi ({\bf r}) \rangle
= n({\bf r}) - \vert \Phi ({\bf r}) \vert ^{2}.
\label{eq18}
\end{equation}
The last two terms in (\ref{eq17}) come from the
$\tilde\psi^\dagger
\tilde\psi$ and $(\tilde\psi^\dagger\tilde\psi)^2$ terms in
(\ref{eq14}).

Calculating the variational derivatives in (\protect\ref{eq8})
using
$F[n,\Phi]$ in (\protect\ref{eq15}),
one finds
\begin{eqnarray}
\frac{\delta F_{s}[n,\Phi]} {\delta n({\bf r})} + v_{H}({\bf r})
+ \frac{\delta F_{xc}[n,\Phi]} {\delta n({\bf r})}
+ v({\bf r})
&=& 0
\nonumber\\
\frac{\delta F_{s}[n,\Phi]} {\delta \Phi ({\bf r})}
+ \frac{\delta F_{xc}[n,\Phi]} {\delta \Phi ({\bf r})}
+ \eta ({\bf r})
&=& 0,
\label{eq19}
\end{eqnarray}
where the Hartree field is defined as
\begin{equation}
v_{H}({\bf r}) \equiv \int d{\bf r}^{\prime} v_{2}
({\bf r} - {\bf r}^{\prime}) n({\bf r}^{\prime}).
\label{eq20}
\end{equation}
Similarly, using (\protect\ref{eq8}) for the auxiliary system
defined
above, one finds
\begin{eqnarray}
\frac{\delta F_{s}[n,\Phi]} {\delta n({\bf r})} + v_{s} ({\bf r})
&=& 0
\nonumber\\
\frac{\delta F_{s}[n,\Phi]} {\delta \Phi ({\bf r})}
+ \eta _{s} ({\bf r})
&=& 0.
\label{eq21}
\end{eqnarray}
Combining the results in (\protect\ref{eq19}) and
(\protect\ref{eq21}), we conclude that the density and order
parameters of the auxiliary system will be identical with the
actual
system if
\begin{eqnarray}
v_{s}({\bf r})
&=&
v({\bf r}) + v_{H}({\bf r}) +
\frac{\delta F_{xc}[n,\Phi]} {\delta n({\bf r})}=v_s[n,\Phi]
\nonumber\\
\eta _{s}({\bf r})
&=&
\eta ({\bf r}) + \frac{\delta F_{xc}[n,\Phi]} {\delta \Phi ({\bf
r})} = \eta _{s} [n,\Phi].
\label{eq22}
\end{eqnarray}
This gives $v_s$ and $\eta_s$ as explicit functionals of $n({\bf
r})$ and
$\Phi({\bf r})$, once we have decided on a specific form for the
functional $F_{xc}[n,\Phi]$.

\section{Bogoliubov gas as auxiliary system}

We now turn to the auxiliary system defined by (\protect\ref{eq9}),
with $\hat V_{s}$ given by (\protect\ref{eq16}). Using
(\protect\ref{eq13}), one finds that
\begin{eqnarray}
\hat H^{s}_{\eta _{{\it s}}, v_{{\it s}}}
&=&
\int d{\bf r} \Phi ^{*} ({\bf r}) [\hat L \Phi ({\bf r}) + \eta
_{s}
({\bf r})] + \int d{\bf r} \eta ^{*}_{s} ({\bf r}) \Phi ({\bf r})
\nonumber\\
&&
+ \int d{\bf r} \tilde \psi ^{\dagger} ({\bf r}) [\hat L \Phi ({\bf
r}) + \eta _{s}({\bf r})]
\nonumber\\
&&
+ \int d{\bf r} [\Phi^{*}({\bf r}) \hat L + \eta ^{*}_{s}
({\bf r})] \tilde \psi ({\bf r})
\nonumber\\
&&
+ \int d{\bf r} \tilde \psi ^{\dagger} ({\bf r}) \hat L \tilde \psi
({\bf r})
\nonumber\\
&&
+ \frac{1}{2} \int d{\bf r} \int d{\bf r}^{\prime} v_{2}
({\bf r} - {\bf r}^{\prime})
\left[ 2 \Phi ^{*} ({\bf r}^{\prime}) \Phi ({\bf r}) \tilde \psi
^{\dagger}
({\bf r}) \tilde \psi ({\bf r}^{\prime}) \right.
\nonumber\\
&&
\left. + \Phi ({\bf r}^{\prime}) \Phi ({\bf r}) \tilde \psi
^{\dagger}
({\bf r}) \tilde \psi ^{\dagger} ({\bf r}^{\prime})
+ \Phi ^{*} ({\bf r}^{\prime}) \Phi ^{*} ({\bf r}) \tilde \psi
({\bf r}) \tilde \psi ({\bf r}^{\prime}) \right],
\label{eq23}
\end{eqnarray}
where the operator $\hat L$ is defined by
\begin{equation}
\hat L \equiv \left(-\frac{\nabla ^{2}} {2m} + v_{s} ({\bf r}) -
\mu
\right).
\label{eq24}
\end{equation}
This Hamiltonian is similar in structure to the one one obtains in
an inhomogeneous weakly interacting Bose gas at $T=0$, which has
been
extensively studied in the literature
\protect\cite{alf-jdw71,alf72}.
We solve it using similar techniques.
In order to diagonalize (\protect\ref{eq23}), we first eliminate
the
terms
linear in $\tilde \psi$ and $\tilde \psi ^{\dagger}$ by requiring
that $\Phi ({\bf r})$ satisfy the equation
\begin{equation}
\hat L \Phi ({\bf r}) + \eta _{s} ({\bf r}) = 0.
\label{eq25}
\end{equation}
Eq. (\protect\ref{eq25}) is a sort of generalized Gross-Pitaevskii
equation for $\Phi$, but now in the context of density functional
theory rather than for a dilute Bose gas. This connection is easily
seen by setting $F_{xc}[n,\Phi]$ in (\protect\ref{eq15}) to zero,
in
which case (\protect\ref{eq22}) simplifies to
\begin{eqnarray}
v_{s} ({\bf r})
&=&
v({\bf r}) + v_{H}({\bf r})
\nonumber\\
\eta _{s} ({\bf r})
&=&
\eta ({\bf r}),
\label{eq26}
\end{eqnarray}
and (\protect\ref{eq25}) reduces to
\begin{equation}
\left[ - \frac{\nabla ^{2}}{2m} + v({\bf r}) - \mu + \int d{\bf r}
^{\prime} v_{2} ({\bf r} - {\bf r} ^{\prime}) n ({\bf r}^{\prime})
\right] \Phi ({\bf r}) + \eta ({\bf r}) = 0.
\label{eq27}
\end{equation}
Setting the external fields $v$ and $\eta$ to zero, we recover the
well-known Gross-Pitaevskii equation \protect\cite{lpp61,alf-jdw71}
for a dilute inhomogeneous gas at $T=0$. In that case, since all
the
atoms are Bose-condensed, the density $n({\bf r})$ in
(\protect\ref{eq27}) can be approximated by $n_{c}({\bf r}) = \vert
\Phi ({\bf r}) \vert ^{2}$, and then (\protect\ref{eq27}) is a
closed
non-linear Schrodinger equation (NLSE) for $\Phi ({\bf r})$.

Assuming that $\Phi ({\bf r})$ satisfies (\protect\ref{eq25}), our
auxiliary system Hamiltonian (\ref{eq23}) reduces to
\begin{eqnarray}
\hat H^{s}_{\eta _{{\it s}}, v_{{\it s}}}
&=&
\int d{\bf r} \eta ^{*}_{s} ({\bf r}) \Phi ({\bf r})
\nonumber\\
&&
+ \int d{\bf r} \tilde \psi ^{\dagger} ({\bf r})
\left[ - \frac{\nabla ^{2}}{2m} + v_{s}({\bf r}) - \mu \right]
\tilde \psi ({\bf r})
\nonumber\\
&&
+ \int d{\bf r} \int d{\bf r} ^{\prime} v_{2} ({\bf r} - {\bf r}
^{\prime}) \Phi ^{*} ({\bf r}^{\prime}) \Phi ({\bf r}) \tilde \psi
^{\dagger} ({\bf r}) \tilde \psi ({\bf r}^{\prime})
\nonumber\\
&&
+ \frac{1}{2} \int d{\bf r} \int d{\bf r} ^{\prime} v_{2} ({\bf r}
- {\bf r} ^{\prime}) \left[ \Phi ({\bf r}^{\prime}) \Phi ({\bf r})
\tilde \psi ^{\dagger} ({\bf r}) \tilde \psi ^{\dagger} ({\bf
r}^{\prime}) + \Phi ^{*} ({\bf r}^{\prime}) \Phi ^{*} ({\bf r})
\tilde \psi ({\bf r}) \tilde \psi ({\bf r}^{\prime}) \right].
\label{eq28}
\end{eqnarray}
The Hartree contribution is contained in $v_{s}({\bf
r})$
[see (\protect\ref{eq22})]. The third line in (\protect\ref{eq28})
gives the exchange term, while the fourth line involves the
anomalous
contributions involving $\tilde \psi ^{\dagger} \tilde \psi
^{\dagger}$ and $\tilde \psi \tilde \psi$ characteristic of a
Bose-condensed system. The quadratic expression given by
(\protect\ref{eq28})
can be diagonalized by the usual Bogoliubov transformation
\protect\cite{alf72}
\begin{eqnarray}
\tilde \psi ({\bf r})
&=&
\sum \limits _{j}\ [u_{j} ({\bf r}) \alpha _{j} - v_{j}^{*}({\bf
r})
\alpha ^{\dagger}_{j}]
\nonumber\\
\tilde \psi ^{\dagger} ({\bf r})
&=&
\sum \limits _{j}\ [u_{j}^{*} ({\bf r}) \alpha _{j}^{\dagger} -
v_{j}({\bf r}) \alpha _{j}],
\label{eq29}
\end{eqnarray}
where the new ``quasiparticle'' operators $\alpha _{j}$ and
$\alpha
_{j}^{\dagger}$ satisfy Bose commutation relations. One finds the
amplitudes $u_{j}({\bf r})$ and $v_{j}({\bf r})$ are given by the
generalized
Bogoliubov coupled equations:
\begin{eqnarray}
\left[ - \frac{\nabla ^{2}} {2m} + v_{s}({\bf r}) - \mu \right]
u_{j}({\bf r})
+ \int d{\bf r}^{\prime} v_{2} ({\bf r} - {\bf r}^{\prime})
\Phi ({\bf r}) \Phi^\ast ({\bf r}^{\prime}) u_{j} ({\bf
r}^{\prime})
&&
\nonumber\\
- \int d{\bf r}^{\prime} v_{2} ({\bf r} - {\bf r}^{\prime})
\Phi ({\bf r}) \Phi ({\bf r}^{\prime}) v_{j} ({\bf r}^{\prime})
&=&
E_{j}u_{j}({\bf r})
\nonumber\\
\left[ - \frac{\nabla ^{2}} {2m} + v_{s}({\bf r}) - \mu \right]
v_{j}({\bf r})
+ \int d{\bf r}^{\prime} v_{2} ({\bf r} - {\bf r}^{\prime})
\Phi^\ast ({\bf r}) \Phi ({\bf r}^{\prime}) v_{j} ({\bf
r}^{\prime})
&&
\nonumber\\
- \int d{\bf r}^{\prime} v_{2} ({\bf r} - {\bf r}^{\prime})
\Phi^\ast ({\bf r}) \Phi^\ast ({\bf r}^{\prime}) u_{j} ({\bf
r}^{\prime})
&=&
-E_{j}v_{j}({\bf r}).
\label{eq30}
\end{eqnarray}
One can prove that the eigenvalues $E_{j}$ are
real and that one must choose solutions such that $E_{j} \geq 0$.
 For details of how (\protect\ref{eq30}) is derived, we
refer
to a similar calculation by Fetter \protect\cite{alf72} for the
case
of a contact interaction $v_{2}({\bf r}) = v_{0} \delta ({\bf r})$.
However, we note that long-range tail of the He-He interatomic
potential is
very important when dealing with the surface region of liquid
Helium
\protect\cite{ag-ss-sub}.

With $u_{j}$ and $v_{j}$ given by the solutions of
(\protect\ref{eq30}), the Hamiltonian in (\protect\ref{eq28}) can
be shown to reduce to
\begin{equation}
\hat H^{s}_{\eta _{{\it s}}, v_{{\it s}}}
= \int d{\bf r} \eta _{s}^{*} ({\bf r}) \Phi
({\bf r})
- \sum \limits _{j} E_{j} \int d{\bf r} \vert v_{j} ({\bf r}) \vert
^{2}
+ \sum \limits _{j} E_{j} \hat \alpha _{j} ^{\dagger} \hat \alpha
_{j},
\label{eq31}
\end{equation}
which describes a non-interacting gas of quasiparticles of energy
$E_j$.
The ground state $\vert \Psi _{s} \rangle$ of this Hamiltonian is
defined by
$\hat \alpha _{j} \vert \Psi _{s} \rangle = 0$. Thus the ground
state expectation value of (\protect\ref{eq31}) is given by
\begin{equation}
E^{s}_{\eta _{{\it s}}, v_{{\it s}}}
= \int d{\bf r} \eta _{s}^{*} ({\bf r}) \Phi
({\bf r}) - \sum \limits _{j} E_{j} \int d{\bf r} \vert v_{j} ({\bf
r}) \vert ^{2}
\label{eq32}
\end{equation}
Using (\ref{eq29}), the non-condensate local density in
(\protect\ref{eq18}) is given by $(T=0)$
\begin{eqnarray}
\tilde n ({\bf r})
&\equiv&
\langle \Psi _{s} \vert \tilde  \psi ^{\dagger} ({\bf r}) \tilde
\psi
({\bf r}) \vert \Psi _{s} \rangle
\nonumber\\
&=&
\sum \limits _{j} \vert v_{j} ({\bf r}) \vert ^{2}.
\label{eq33}
\end{eqnarray}
Inserting (\protect\ref{eq31}) into (\protect\ref{eq12}) and using
(\protect\ref{eq25}) and (\protect\ref{eq18}), one finds after a
little algebra that
\begin{equation}
F_{s}[n,\Phi] =
\int d{\bf r} \Phi ^{*} ({\bf r}) \left[ - \frac{\nabla ^{2}} {2m}
- \mu \right] \Phi ({\bf r}) - \sum \limits _{j} \int d{\bf r}
\vert
v_{j} ({\bf r}) \vert ^{2} \left[ E_{j} + v_{s} ({\bf r}) \right]\
{}.
\label{eq34}
\end{equation}
We recall that the Hartree contribution was separated out in
(\protect\ref{eq15}) and consequently it is not contained in
$\hat H^{s}_{\eta _{{\it s}}, v_{{\it s}}}$
defined in (\protect\ref{eq23}). Thus it is not included
in the energy eigenvalues $E_{j}$ given by (\protect\ref{eq30}),
but
rather appears as a separate contribution from the diagonal
potential
$v_{s}
({\bf r})$ in (\protect\ref{eq22}).

The key feature of the above results  is that one can
have a
depletion of the condensate, as shown by the finite value of
$\tilde
n ({\bf r})$ in (\protect\ref{eq33}). Thus the auxiliary system
defined by (\protect\ref{eq23}) and (\protect\ref{eq22}) can be
used
to find both $\Phi ({\bf r})$ and $n({\bf r})$  [using (\ref{eq25})
and (\ref{eq30})]  even when $n_{c}
({\bf
r}) \equiv \vert \Phi ({\bf r}) \vert ^{2}$ and $n({\bf r})$ are
quite
different (as in superfluid $^4$He).

If we had chosen a non-interacting Bose gas as our KS reference
system [{\it i.e.,}\ set $\hat V_s =0$ in (\ref{eq9})], the last
term in (\ref{eq23}) would be absent.  The linear terms in
$\tilde\psi$ and $\tilde\psi^\dagger$ can be eliminated as before
by requiring that $\Phi({\bf r})$ satisfy (\ref{eq25}). Then $\hat
H^s_{\eta_s, v_s}$ is easily diagonalized, with $u_j({\bf r})$ given
by the solution
\begin{equation}
\left (-{\nabla^2\over 2m} +v_s({\bf r})-\mu\right)u_j({\bf r}) =
E_ju_j({\bf
r})
\label{neweq35}
\end{equation}
and $v_j({\bf r})=0$.  It immediately follows from (\ref{eq33})
that such a KS reference system leads to no depletion, with
$\tilde n({\bf r})=0$.
This means that using a free Bose gas as a reference system
inevitably leads to $n_c({\bf r})\equiv \vert\Phi({\bf r})\vert^2$
and $n({\bf r})$ being identical, no matter what we choose for the
functional $F_{xc}[n,\Phi]$. This apparent ``insufficiency'' of
the non-interacting Bose gas is somewhat surprising and deserves
further study.  It seems to
be associated with the well-known fact that a weakly interacting
Bose-condensed gas has qualitatively different properties than an
ideal Bose gas.

It is convenient at this point
to summarize the various steps in the KS procedure:

\begin{enumerate}
\item [\rm (1)] One chooses some approximation for the
``exchange-correlation'' functional $F_{xc}[n,\Phi]$ defined in
(\protect\ref{eq15}), giving it as an explicit functional of
the local quantities $n({\bf r})$ and $\Phi ({\bf r})$.
This is the big step containing the ``physics'', and has not been
addressed in the present paper.
\item [\rm (2)] The potentials $v_{s}({\bf r})$ and $\eta
_{s}({\bf r})$ are then computed using (\protect\ref{eq22}) and
given as functionals of $n({\bf r})$ and $\Phi ({\bf r})$.
\item [\rm (3)] Evaluating $v_{s}$ and $\eta _{s}$ using an assumed
(or trial) value of $n({\bf r})$, the GP-type equation
(\protect\ref{eq25})
is solved for $\Phi ({\bf r})$, i.e.,
\begin{equation}
\left[ - \frac{\nabla ^{2}}{2m} + v_{s}[n,\Phi]  - \mu \right]
\Phi ({\bf r}) + \eta _{s} [n,\Phi] = 0.
\label{eq35}
\end{equation}
\item [\rm (4)] Evaluating $v_{s}[n,\Phi]$ using the trial
$n({\bf r})$ and the solution for $\Phi ({\bf r})$ given in step
(3),
the  generalized Bogoliubov equations in (\protect\ref{eq30}) can
be solved to
determine
$E_{j}, u_{j} ({\bf r})$ and $v_{j} ({\bf r})$.
\item [\rm (5)] With these results, one can calculate the
non-condensate local density $\tilde n ({\bf
r})$ in (\protect\ref{eq33}) and hence finally obtain $n({\bf r})$
from
(\protect\ref{eq18}), namely
\begin{equation}
n({\bf r}) = \vert \Phi ({\bf r}) \vert ^{2}+\sum_{j}|v_{j}({\bf
r})|^2.
\label{eq36}
\end{equation}
Using this new expression for $n({\bf r})$, one can go back to step
(3) and
repeat the procedure, until self-consistent values of $\Phi ({\bf
r})$ and
$n({\bf r})$ are obtained.
\item [\rm (6)] These values of $n({\bf r})$ and $\Phi ({\bf r})$
are
then inserted into the energy functional of the actual Bose system
of
interest, as given by (\protect\ref{eq7}) and (\protect\ref{eq15}).
\end{enumerate}

\section{Conclusions}

As in the case of interacting Fermi systems
\protect\cite{wk-ljs65,lno-ekg-wk88}, it is important to
emphasize that while the auxiliary Bose system described in Section
IV
corresponds to a dilute, weakly interacting Bose gas, we are only
using it to find the local density $n({\bf r})$ and local order
parameter
$\Phi ({\bf r})$ of a Bose-condensed liquid. In particular, the
quasiparticle
excitations which are described by the generalized Bogoliubov
equations of
motion
in (\protect\ref{eq30}) have, in general, no ``direct'' physical
significance in
the true system.

The above procedure gives a well-defined scheme to find (at $T=0$)
the energy, local density and local condensate density in
superfluid
$^4$He with a free surface, once one has chosen some explicit
approximation for the exchange-correlation functional $F_{xc}
[n,\Phi]$ in (\protect\ref{eq15}). As in the case of normal and
superconducting metals, a good approximation for this functional is
the key to
obtaining
reasonable results using the density functional formalism. We hope
to
discuss this problem elsewhere. In the case of superfluid $^4$He,
there has
been considerable work \protect\cite{ce-wfs-etc75} on constructing
functionals which only depend on the total local density $n({\bf
r})$, in which one tries to build in known experimental information
(compressibility, ground state energy, surface tension, etc.). In
developing the equivalent approximations for use in our new
formalism, the
first thing we need to understand better is the ground-state energy
of {\it bulk} liquid $^4$He as a function of the density $n$ and
the
condensate density $n_{0}$. More Monte Carlo calculations would be
very
useful.

One immediate implication of the present theory is contained in the
first term of (\protect\ref{eq34}), which can be rewritten in the
form (for clarity, we insert $\hbar$)
\begin{equation}
\int d{\bf r} \Phi ^{*} ({\bf r}) \left[ - \frac{\hbar ^{2} \nabla
^{2}} {2m} \right] \Phi ({\bf r})
= \int d{\bf r} \frac{\hbar ^{2}}{2m} \vert \nabla \sqrt{n_{c}({\bf
r})} \vert ^{2} + \frac{1}{2} \int d{\bf r} mn_{c} ({\bf r}) {\bf
v}^{2}_{s}({\bf r}),
\label{eq37}
\end{equation}
where we have used $\Phi ({\bf r}) = \sqrt{n_{c}({\bf r})}
e^{iS({\bf
r})}$ and $m{\bf v}_{s}({\bf r}) \equiv \hbar\nabla S({\bf r})$ is
the
local
superfluid velocity (in the ground state, we can set ${\bf
v}_{s}({\bf r}) = 0$). In contrast with the first term in
(\protect\ref{eq37}), currently available density functional
theories
\protect\cite{ce-wfs-etc75} always start with a term of the form
\begin{equation}
\int d{\bf r} \frac{\hbar ^{2}}{2m} \vert \nabla \sqrt{n({\bf
r})}\vert ^{2},
\label{eq38}
\end{equation}
involving the {\it total} local density $n({\bf r})$. This
corresponds to the kinetic energy functional of an inhomogeneous
non-interacting Bose gas constrained to have the correct local
density
of the Bose liquid under consideration. We believe that the first
term in (\protect\ref{eq37}) has a more natural as well as more
sound
theoretical basis when dealing with Bose-condensed fluids, and
should be used in developing improved density functional theories
of
superfluid $^4$He \protect\cite{ag-ss-sub}.

At the present time, the only detailed discussion of the properties
of a spatially inhomogeneous Bose-condensed system has been for a
weakly interacting atomic gas trapped in an external potential well
\cite{newref23}.  In such systems \cite{newref22}, the condensate
density $n_c({\bf r})$ is strongly peaked at the center of the trap
(due to the macroscopic occupation of the lowest quantum state) and
has a very different spatial dependence from the non-condensate
density $\tilde n({\bf r})$. Our density functional formalism
(extended to finite temperatures) should be useful in calculating
the properties of
such trapped Bose-condensed gases.
\section*{Acknowledgements}

I would like to acknowledge many stimulating discussions with
Sandro
Stringari, which led to this work. I also thank Hardy Gross and the
referee for
useful comments. This work was done while on sabbatical at the
University of Trento, which provided financial support and a
congenial atmosphere. I also thank NSERC for a research grant.

\end{document}